\documentclass[a4paper,twoside,12pt]{article}
\input{obsjkt.sty}

\newcommand{\Msun}{\ensuremath{~{\rm M}_\odot}}                   
\newcommand{\Rsun}{\ensuremath{~{\rm R}_\odot}}                   
\newcommand{\Lsun}{\ensuremath{~{\rm L}_\odot}}                   
\newcommand{\rhosun}{\ensuremath{~\rho_\odot}}                    
\newcommand{\Teff}{\ensuremath{T_{\rm eff}}}                      
\newcommand{\FeH}{\ensuremath{\rm [Fe/H]}}                        
\newcommand{\EBV}{\ensuremath{E(B\!-\!V)}}                        
\renewcommand{\kms}{~km~s$^{-1}$}                                 
\newcommand{\gaia}{\textit{Gaia}}                                 
\newcommand{\targ}{BK~Peg}
\newcommand{\targfull}{BK~Pegasi}

\newcommand{\Msunnom}{\hbox{$\mathcal{M}^{\rm N}_\odot$}}
\newcommand{\Rsunnom}{\hbox{$\mathcal{R}^{\rm N}_\odot$}}
\newcommand{\Lsunnom}{\hbox{$\mathcal{L}^{\rm N}_\odot$}}

\usepackage{rotating}
\usepackage{pdflscape}

\begin{document}

\OBSheader{Rediscussion of eclipsing binaries: \targ}{A.C. Kutluay \& J. Southworth}{2026 June}

\OBStitle{Rediscussion of eclipsing binaries. XXX. \\ The slightly evolved F-type system BK Pegasi}

\OBSauth{Ahmet Cem Kutluay and John Southworth}

\OBSinstone{Astrophysics Group, Keele University, Staffordshire, ST5 5BG, UK}

\OBSabstract{\targ\ is a double-lined detached eclipsing binary containing two late-F stars in an orbit with small eccentricity. We use light curves from the Transiting Exoplanet Survey Satellite (TESS) and spectroscopic measurements from previous studies to measure the physical properties of the companions to a high precision. We obtain masses of $1.411 \pm 0.004$\Msun\ and $1.254 \pm 0.004$\Msun, and radii of $1.990 \pm 0.004$\Rsun\ and $1.460 \pm 0.004$\Rsun, which are among the most precise measurements made for these quantities in normal stars. These properties match theoretical stellar evolutionary models for a solar chemical composition and an age of 2.65~Gyr. We also present an updated ephemeris of the system, as a result of our TESS measurements and a collection of mid-eclipse times from previous studies.}


\section*{Introduction}

This study is part of the ongoing series \cite{Me20obs} in which known detached eclipsing binary systems (dEBs) are re-analysed based on new photometric data, primarily obtained with the NASA Transiting Exoplanet Survey Satellite \cite{Ricker+15jatis} (TESS). Our main objective is to exploit space-based observations \cite{Me21univ} to refine the measurements of the stellar components’ properties and to incorporate these systems into the Detached Eclipsing Binary Catalogue \cite{Me15aspc} (DEBCat\footnote{\texttt{https://www.astro.keele.ac.uk/jkt/debcat/}}).

In this work, we present a study of \targfull\  (Table~\ref{tab:object_properties}), a dEB composed of two late-F stars in a slightly eccentric orbit. The system has the unusual characteristic that its more massive and larger primary component (hereafter star~A) has a slightly lower effective temperature than the secondary (star~B). This is a consequence of primary's ongoing evolution toward the sub-giant stage on the HR diagram\cite{clausen1990}. 

Hoffmeister\cite{hoffmeister1931} originally identified its eclipsing character and classified the system as an Algol-type variable. The first determination of the orbital period was carried out by Lause\cite{lause1935,lause1937}, yielding a value of 2.745 days. Later, Popper \& Dumont\cite{popperdumont1977} revealed that light curves of \targ\ contain two eclipse minima of almost identical depth (Fig.~\ref{fig:tess}). They corrected the orbital period value to 5.49 days, which was later refined by Clausen et al\cite{clausen2010} (hereafter CL10).

Initial estimates of the absolute dimensions were reported by Popper\cite{popper1980} in his review of stellar masses, followed by spectroscopic analyses and refined determinations of the absolute parameters in a subsequent study\cite{popper1983}. According to these two studies, the masses (1.43\Msun\ and 1.28\Msun) and luminosities (4.68\Lsun\ and 3.09\Lsun) of the components differ significantly. Following this study, Demircan et al.\cite{demircan1994} obtained $UBV$ light curves of the system and refined the absolute dimensions, using the radial velocities (RVs) given by Popper\cite{popper1983}.

In addition, Demircan et al.\ discussed the evolutionary status of the system using the mass–radius $(M-R)$, mass–luminosity $(M-L)$, and temperature–luminosity $(T-L)$ planes based on evolutionary models\cite{schaller1992}, claiming that the components are still in the core hydrogen-burning phase and best represented by high-metallicity models with an age of about $3.3$ Gyr. More recently, CL10 presented high-precision absolute dimensions and spectroscopic chemical abundances for BK Peg, showing that the components have evolved to the upper half of the main-sequence band. Their comparison with scaled-solar evolutionary models suggested slightly younger ages ($\sim$2.5 to 2.8 Gyr), with indications that the amount of convective core overshooting may affect the inferred evolutionary status.

\begin{table}[t]
    \caption{\em Basic information on \targfull. 
    The $BV$ magnitudes are each the mean of 94 individual measurements \cite{hog2000} distributed approximately randomly in orbital phase. The $JHK_s$ magnitudes are from 2MASS \cite{cutri2003} and were obtained at an orbital phase of 0.89.}
    \centering
    \begin{tabular}{lll}
    {\em Property} & {\em Value}  & {\em Reference} \\
    Right Ascension (J2000) & 23 47 08.26 & \citenum{gaiaedr3}\\
    Declination (J2000)&  +26 33 59.97 & \citenum{gaiaedr3} \\
    Bonner Durchmusterung designation & BD+25 5003 & \citenum{argelander1903}\\
    Tycho designation & TYC 2254-2563-1 & \citenum{hog2000}\\
    Gaia DR3 designation & 2852979962499356288 & \citenum{gaiaedr3}\\
    Gaia DR3 parallax (mas) & $3.2643 \pm 0.0177$ & \citenum{gaiaedr3} \\
    TESS Input Catalogue designation & TIC 269747005 & \citenum{stassun2019}\\
    $B$ magnitude & 10.46 & \citenum{hog2000} \\
    $V$ magnitude & 10.04 & \citenum{hog2000}\\
    $G$ magnitude & 9.835 & \citenum{gaiaedr3}\\
    $J$ magnitude & 8.892 & \citenum{cutri2003}\\
    $H$ magnitude & 8.643 & \citenum{cutri2003}\\
    $K_s$ magnitude & 8.611 & \citenum{cutri2003}\\
    Spectral type & F8~V + F7~V & \citenum{clausen2010}\\ 
    \end{tabular}
    \label{tab:object_properties}
\end{table}

\section*{Photometric observations}

\targ~ has been observed by TESS in two sectors (Sector 57 in October 2022 and Sector 84 in October 2024). For Sector 57, short-cadence data with 120~s sampling were available, whereas only full-frame images (FFIs) are available for Sector 84. Therefore, our analysis is based solely on the 120~s cadence data from Sector 57. The data were retrieved from the NASA Mikulski Archive for Space Telescopes (MAST\footnote{\texttt{https://mast.stsci.edu/portal/Mashup/Clients/Mast/Portal.html}}) via the {\sc lightkurve} package \cite{Lightkurve18}.

\begin{figure}[t] 
\centering 
\includegraphics[width=\textwidth]{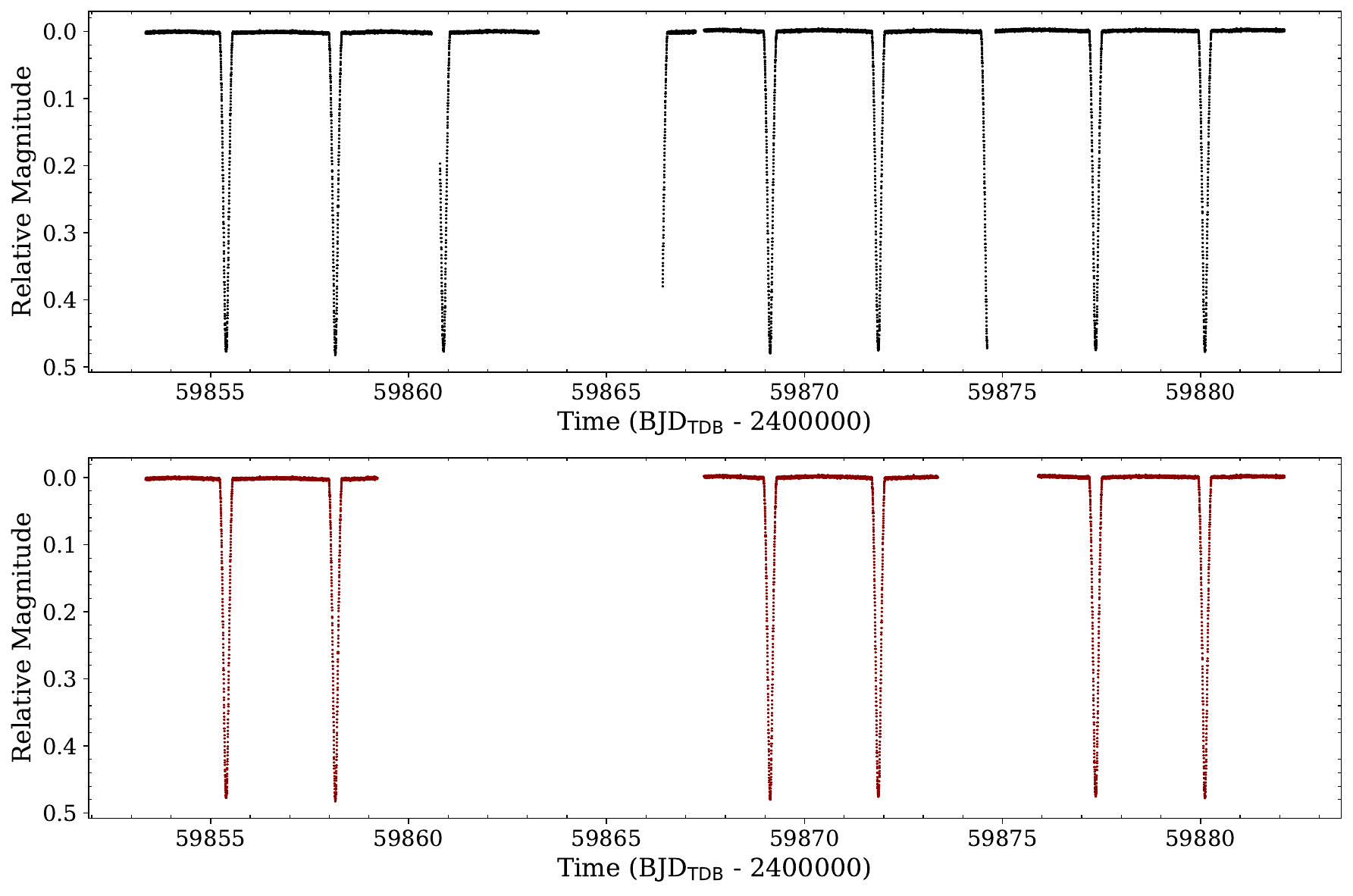} \\
\caption{\label{fig:tess} Top: TESS sector 57 photometry of \targ. The flux measurements have been converted to magnitude units, and the median was subtracted. Bottom: trimmed light curves of Sector 57 for further analysis.}
\end{figure}

For our analysis, we utilized the simple aperture photometry (SAP) light curves produced by the SPOC data reduction pipeline \cite{Jenkins+16spie}, excluding data points flagged as low-quality using the {\sc lightkurve} ``hard'' quality flag. In Sector 57, the data are not fully continuous and contain gaps. For this reason, portions of the light curve where the eclipses were poorly sampled or affected by gaps were manually rejected and only the remaining data were retained for further analysis. The entire sector is shown in the top panel of Fig.~\ref{fig:tess}, while the manually trimmed segments are displayed in the bottom panel. The remaining data were converted into differential magnitudes and the median magnitude was subtracted for convenience.

\begin{figure}[t] 
\centering 
\includegraphics[width=\textwidth]{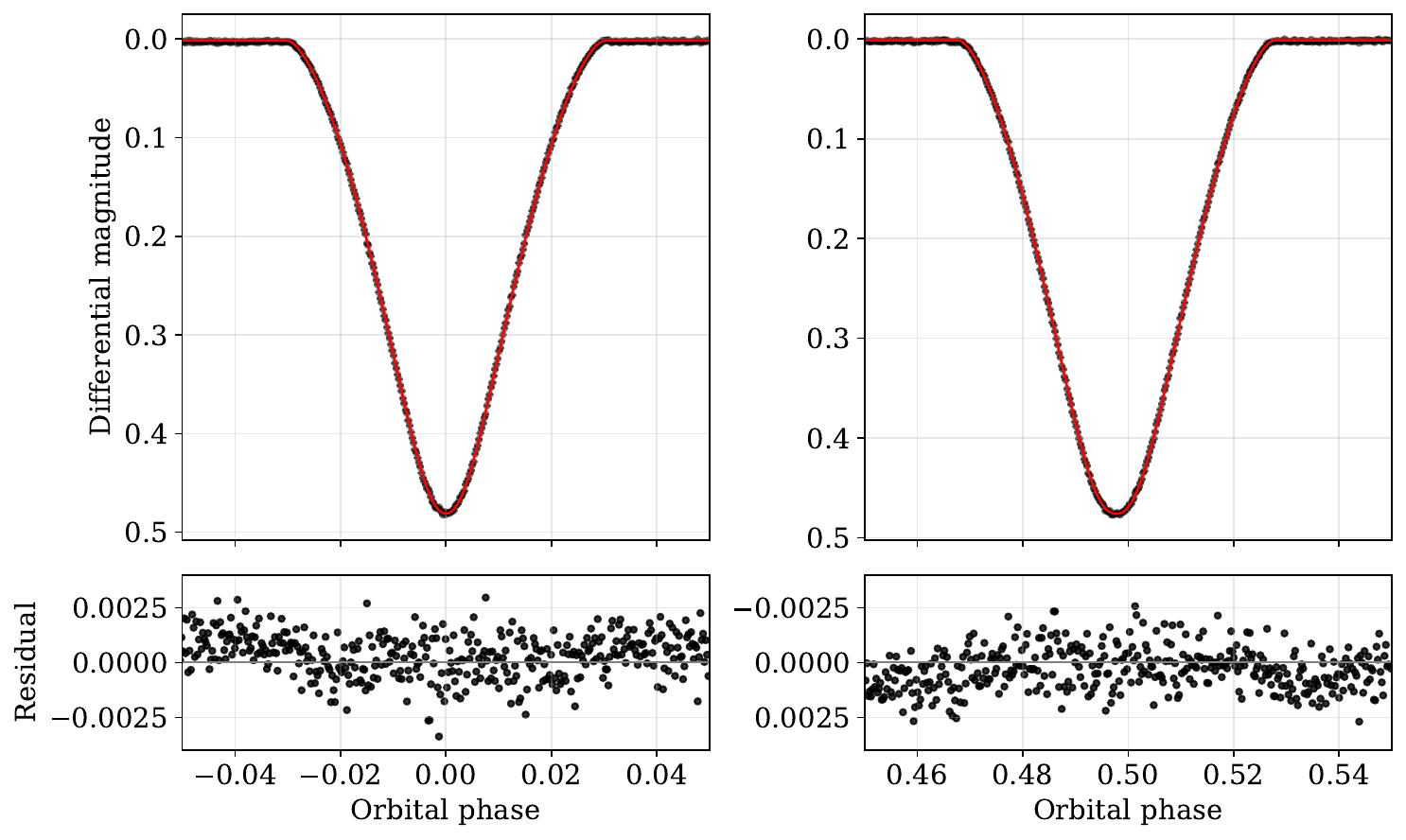} \\
\caption{Best-fit model to the TESS Sector 57 light curves of {\targ} for the primary eclipse (left panels) and secondary eclipse (right panels), obtained using \textsc{jktebop}. The observed data are shown as black points, and the model fit is plotted as a red solid line. The residuals are displayed on an enlarged scale in the lower panels.}
\label{fig:lc} 
\end{figure}

\section*{Light curve analysis}

As shown in Fig.~\ref{fig:lc}, {\targ} is a well-detached binary system with two eclipses of nearly equal depth of approximately 0.5 mag. This makes the system a suitable target for modelling with the {\sc jktebop}\footnote{\texttt{http://www.astro.keele.ac.uk/jkt/codes/jktebop.html}} code\cite{Me++04mn2,Me13aa}. Due to the orbital period of the system ($\sim$ 5.5 days) and the duration of the TESS sector ($\sim27$ days), only five orbital cycles were observed.
For each eclipse, all data during the event were extracted from the TESS light curve, along with additional segments immediately before and after the eclipse. These additional segments cover at least 30 minutes and are useful for setting the out-of-eclipse brightness of the system. Each eclipse was then normalised to zero differential magnitude by fitting and subtracting a straight line to the out-of-eclipse data, thereby removing slow instrumental or astrophysical trends. Subsequently, second-order polynomial fits were independently applied to the out-of-eclipse regions in three separate portions of the TESS light curve to further suppress low-frequency variations and minimise the residuals.

The fitted parameters were the fractional radii of the stars ($r_{\rm A}$ and $r_{\rm B}$), the central surface brightness ratio ($J$), third light ($L_3$), orbital inclination ($i$), eccentricity ($e$), argument of periastron ($\omega$), orbital period ($P$), and a reference time of primary minimum ($T_0$). The fractional radii were expressed as their sum ($r_{\rm A}+r_{\rm B}$) and ratio ($k = {r_{\rm B}}/{r_{\rm A}}$), and the orbital shape parameters as the combinations $e\cos\omega$ and $e\sin\omega$ to decrease correlations between parameters. Limb darkening (LD) was implemented using the power-2 law\cite{hestroffer1997, maxted2018, southworth2023ld}, with the linear coefficient ($c$) fitted and the non-linear coefficient ($\alpha$) fixed at a theoretical value\cite{claretsouthworth2022,claretsouthworth2023}. The same LD coefficients were used for both stars due to their almost identical temperature. To ensure that the adopted photometric uncertainties accurately represent the observed scatter, the TESS flux measurement errors were iteratively scaled until the reduced chi-square of the fit, $\chi^2_\nu$, reached unity. An example fit is shown in Fig.~\ref{fig:lc}. Our fitted light ratio is in excellent agreement with the spectroscopically-determined value of $0.57 \pm 0.04$ from CL10.

\begin{table}
\centering
\caption{\em \label{tab:jktebop_lc} Photometric parameters of \targ\ measured using {\sc jktebop} from the TESS light curves. The error bars are $1\sigma$ standard errors obtained from the Monte Carlo simulations applied to Sector $57$ data. Also, we provide the results from the $y$-filter light-curves in CL10 for comparison.}
\begin{tabular}{lcc}
{\em Parameter} & {\em Value} & CL10($y$) \\[3pt]
{\it Fitted parameters:} \\
Orbital inclination ($^\circ$) &  $88.14 \pm 0.17$ & $88.02 \pm 0.05$\\
Sum of the fractional radii & $0.19001 \pm 0.00009$ & $0.1898 \pm 0.0005$\\
Ratio of the radii &  $0.733 \pm 0.094$ & $0.7379\pm0.0050$\\
Central surface brightness ratio & $1.0432\pm0.0038$ & $1.0444 \pm 0.0031$ \\
Third light & $0.01 \pm 0.10$ & 0.0 (fixed) \\
$e\cos\omega$ & $-0.00356 \pm 0.00001$ & $-0.00364\pm 0.00006$ \\
$e\sin\omega$ & $-0.0010 \pm 0.0022$ & $0.00283 \pm 0.00181$\\
LD coefficient $c$ & $0.614 \pm 0.009$ & \\
LD coefficient $\alpha$ & 0.515 (fixed) & \\
{\it Derived parameters:} \\
Fractional radius of star~A & $0.1096 \pm 0.0002$  & 0.1092 \\
Fractional radius of star~B & $0.0804 \pm 0.0002$  & 0.0806\\
Light ratio $\ell_{\rm B}/\ell_{\rm A}$ & $0.5608 \pm 0.0009$ & 0.5670 \\[3pt]
Orbital eccentricity & $0.00368 \pm 0.00004$  & 0.0046 \\
Argument of periastron ($^\circ$) & $195.15 \pm 0.12$ & 142.1 \\
\end{tabular}
\end{table} 

We calculated uncertainties in the fitted parameter values using 10,000 Monte Carlo simulations. The results of this analysis are provided in Table~\ref{tab:jktebop_lc} with a comparison to the results from $y$-band light curves of CL10. We have increased the error bars for $r_{\rm A}$ and $r_{\rm B}$ to 0.2\% because it has not yet been demonstrated that light curve models are reliable beyond this level of precision\cite{maxted2020}. Overall, our results are in good agreement with those in CL10. The only noticeable difference is found in the eccentricity, which is lower in our solution by approximately 20\%. Although the formal errors for $e$ and $\omega$ in our table are very small, they primarily reflect the statistical precision of the fit rather than physical certainty. As a consequence of the very small eccentricity, the argument of periastron $\omega$ is highly subject to model degeneracies, and such an apparent difference between our value of $\omega$ and that reported by CL10 is likely dominated by model degeneracies rather than a physically significant change in the orbit. We leave the discussion of apsidal motion to future study, emphasizing the necessity of ongoing measurements of times of primary and secondary eclipse.

\section*{Orbital ephemeris}

This work includes only one sector of TESS data, which consists of three intervals of photometric observations (Fig.~\ref{fig:tess}). We modelled each interval individually with {\sc jktebop} and determined the mid-times of primary eclipses. As these three intervals span only 25~d, we collected three times of primary minimum from previous studies\cite{demircan1994,ak2003} and eight times of minimum from the VarAstro portal of Variable Star and Exoplanet Section of Czech Astronomical Society\footnote{\texttt{https://var.astro.cz/en/}}. All primary minimum times are presented in Table~\ref{tab:tmin}.

We fitted a linear ephemeris to these minimum times with Python's {\sc scipy} package\cite{Virtanen+20scipy}, obtaining

\begin{equation}
\mbox{Min~I} = {\rm BJD}_{\rm TDB}~ 2450706.46975(20) +  5.48991130 (12) E
\end{equation}

\begin{table} \centering
\caption{\em Times of mid-eclipse for \targ\ and their residuals versus the fitted ephemeris. \label{tab:tmin}}
\setlength{\tabcolsep}{10pt}
\begin{tabular}{rrrrr}
{\em Orbital} & {\em Eclipse time}  & {\em Uncertainty} & {\em Residual} & {\em Source}   \\
{\em cycle}   & {\em (BJD$_{TDB}$)} & {\em (d)}         & {\em (d)}      &  \\[3pt]               
-329.0 & 2448900.28904 & 0.00056   & 0.00106 & Demircan et al.\cite{demircan1994} \\
0.0    & 2450706.47007 & 0.00030   & 0.00106 & Ak et al.\cite{ak2003} \\
63.0   & 2451052.33246 & 0.00070   & -0.00100 & Ak et al.\cite{ak2003} \\
751.0  & 2454829.39325 & 0.00070   & 0.00037 & Hübscher et al.\cite{hubscher2010} \\
931.0  & 2455817.57785 & 0.00001   & 0.00081 &  varAstro \\
990.0  & 2456141.48076 & 0.00001   & -0.00107 & varAstro \\
1004.0 & 2456218.34064 & 0.00001   & 0.00003 & varAstro \\
1398.0 & 2458381.36621 & 0.00001   & 0.00029 & varAstro \\
1531.0 & 2459111.52508 & 0.00001   & 0.00087 & varAstro \\
1543.0 & 2459177.40241 & 0.00001   & -0.00073 & varAstro \\
1667.0 & 2459858.15185 & 0.00002   & -0.00038 & TESS Sector 57 \\
1669.0 & 2459869.13184 & 0.00008   & -0.00021 & TESS Sector 57 \\
1671.0 & 2459880.11153 & 0.00002   & -0.00035 & TESS Sector 57 \\
\end{tabular}
\end{table}

\begin{figure} 
\centering 
\includegraphics[width=\textwidth]{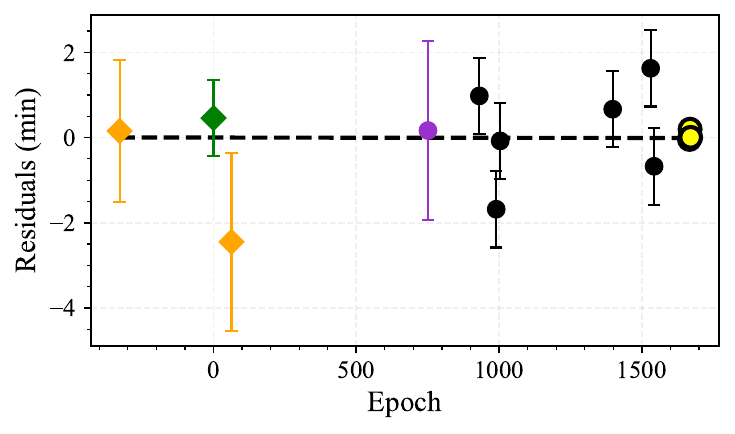}
\caption{Residuals of the primary minimum times from Demircan et al.\cite{demircan1994} (green), Ak et al.\cite{ak2003} (orange), Hübscher et al.\cite{hubscher2010} (purple),
VarAstro (black) and TESS (yellow), as listed in Table~\ref{tab:tmin}. Dashed black line represents the residuals of zero according to the linear ephemeris. Diamonds represent the photoelectric observations, while circles denote the CCD observations.}
\label{fig:ocdiagram} 
\end{figure}

\noindent where $E$ corresponds to the number of cycles since a reference time of minimum, and the numbers in parentheses show the uncertainties in the final significant figure of the corresponding measurement. The root-mean-square of the residuals is 61~s which is higher than most of the errorbars suggest, and the reduced $\chi^2$ is $\chi_\nu^2 = 4.32$. The uncertainties in the ephemeris have been multiplied by $\sqrt{\chi_\nu^2}$ to account for this high $\chi_\nu^2$.

Popper \& Etzel\cite{popperetzel1981} and Demircan et al.\cite{demircan1994} also noted the variability of individual light curves and suggested that one of the components of \targ\ might be a pulsating star. We checked the TESS data for the presence of pulsations using the {\sc period04} code\cite{period04}. We found no significant pulsation signal, to a limit of 0.2\,mmag, but we see hints of starspot activity. Such activity could affect the fit of the light curve and the times of minimum obtained from it.

Due to its slight eccentricity, \targ\ is expected to experience slow apsidal motion which could also increase the scatter when attempting to fit a linear ephemeris. However, we see no hint of this effect in Fig.\,\ref{fig:ocdiagram}. We leave an analysis of apsidal motion to future work.

\begin{figure}
\centering 
\includegraphics[width=\textwidth]{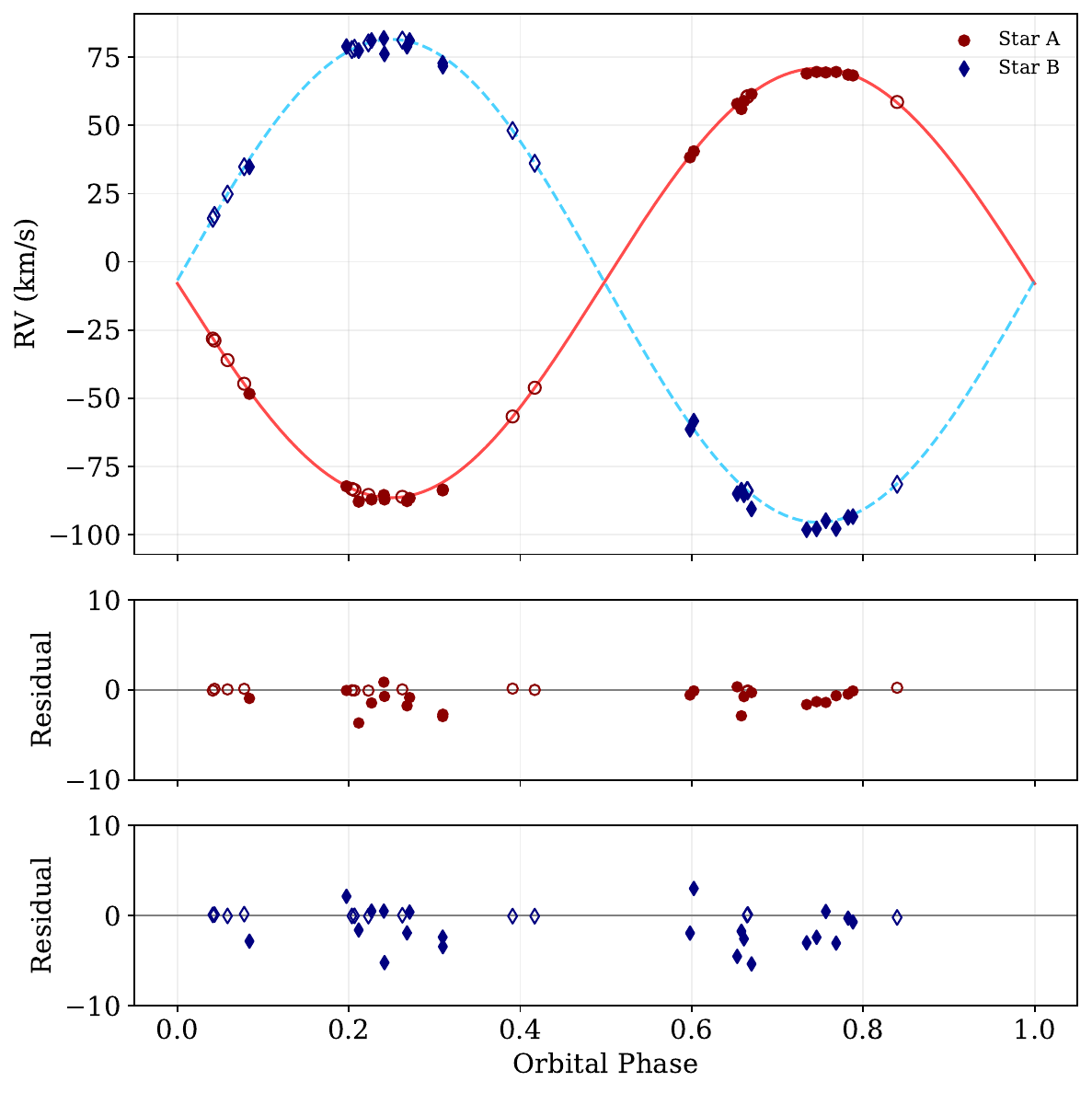} \\
\caption{\label{fig:rv} RVs of \targ\ compared to the best fit from 
{\sc jktebop} (solid blue lines). The RVs for star~A are shown with dark red colour, and for star~B with blue colour. The residuals are given in the lower panels separately for the two components. RVs from Popper\cite{popper1983} are shown with closed circles and diamonds, and those from CL10 with open circles and diamonds.} \end{figure}

\section*{Radial velocity analysis}

\targ\ has been the subject of two previous spectroscopic studies: Popper\cite{popper1983} and CL10. The former obtained 22 RV measurements for each star from photographic spectra obtained at Lick Observatory, whereas the latter provided 13 RV measurements per star from the FIES \'echelle spectrograph at the Nordic Optical Telescope\cite{fies2014}. CL10 compared their results with Popper and concluded that they are in a good agreement within errors, although the systemic velocities ($V_{\gamma,A},V_{\gamma,B}$) differ by about 1 km s$^{-1}$.   

In this study, four different analyses were performed based on the RV data source: Popper only, CL10 only, the combined RV dataset with separate $V_{\gamma}$ values for the two stars, and the combined RV dataset with the same $V_{\gamma}$ for the stars. In each case, we fitted the RV data using {\sc jktebop} with a fixed $P$ but allowing for a shift in $T_0$. The other fitted parameters were $K_{\rm A}$, $K_{\rm B}$, $V_{\rm\gamma,A}$ and $V_{\rm\gamma,B}$. We kept $e\cos \omega$ and $e\sin \omega$ fixed at the photometric values provided at Table~\ref{tab:jktebop_lc}. We present our results in Table~\ref{tab:spec_orbit} and the phase-folded RV measurements from both sources, along with their model residuals, are shown in Fig.~\ref{fig:rv}. In all instances, the error bars were calculated from 5000 Monte Carlo simulations. We conclude that all solutions are in agreement with the previous results of Popper and CL10 within the error bars. For our final result we took the average of the $K_{\rm A}$ and $K_{\rm B}$ values from our two fits to the combined data, but added the difference between the values for the two fits to the quoted uncertainties. This provided an additional uncertainty on the results stemming from our choice of which model to apply.

\section*{Physical properties and distance to \targ}

Physical properties of \targ\ were calculated using the {\sc jktabsdim} code \cite{jktabsdim} with the photometric properties from Table~\ref{tab:jktebop_lc} and the spectroscopic properties from Table~\ref{tab:spec_orbit}. We adopted the effective temperatures $T_{\rm eff,A} = 6265 \pm 85 {\rm ~K}$ and $T_{\rm eff,B} = 6320 \pm 90 {\rm ~K}$ from CL10. We present our results in Table~\ref{tab:absdim}. We have achieved a precision of 0.3\% in both mass and radius, and our measurements also agree with those from CL10 and Popper. 

\begin{landscape}
\begin{table}
\centering
\caption{\em Spectroscopic orbits for \targ\ from the literature and from the current work. All quantities are given in km~s$^{-1}$. The error bars are $1\sigma$ standard errors obtained from the Monte Carlo simulations.}
\label{tab:spec_orbit}
\setlength{\tabcolsep}{10pt}
\begin{tabular}{lcccccc}
{\em Source} 
& $K_{\rm A}$  & $K_{\rm B}$  & $V_{\gamma,{\rm A}}$ & $V_{\gamma,{\rm B}}$ & $\sigma_{\rm A}$ & $\sigma_{\rm B}$ \\[6pt]
Popper \cite{popper1983}                 & $79.14 \pm 0.33$ & $88.86 \pm 0.56$ & $-8.51 \pm 0.30$ & $-8.35 \pm 0.51$ & $1.36$ & $2.30$ \\
CL10                                     & $78.77 \pm 0.11$ & $88.59 \pm 0.21$ & $-7.39 \pm 0.08$ & $-7.20 \pm 0.15$ & $0.26$ & $0.50$ \\[3pt]
This work (Popper RVs)                   & $79.02 \pm 0.26$ & $88.86 \pm 0.49$ & $-8.59 \pm 0.24$ & $-8.93 \pm 0.45$ & $1.08$ & $2.13$ \\
This work (CL10 RVs)                     & $78.70 \pm 0.04$ & $88.51 \pm 0.04$ & $-7.47 \pm 0.03$ & $-7.29 \pm 0.03$ & $0.11$ & $0.09$ \\
This work (all RVs, separate $V_\gamma$)   & $78.70 \pm 0.06$ & $88.51 \pm 0.05$ & $-7.49 \pm 0.04$ & $-7.30 \pm 0.03$ & $1.00$ & $2.14$ \\
This work (all RVs, common $V_\gamma$) & $78.75 \pm 0.06$ & $88.55 \pm 0.04$ & $-7.37 \pm 0.03$ & $-7.37 \pm 0.03$ & $1.31$ & $2.11$\\[3pt]
{Adopted spectroscopic orbit} & $78.72 \pm 0.11$ & $88.53 \pm 0.09$ & $-7.43 \pm 0.16$ & $-7.34 \pm 0.10$  &  1.00 & 2.14 \\
\end{tabular}
\end{table}
\end{landscape}

We measured the distance to \targ\ using the $BV$ magnitudes from Tycho\cite{hog2000}, $JHK_s$ magnitudes from 2MASS\cite{cutri2003} corrected onto the Johnson system, and the surface brightness calibrations of Kervella et al.\cite{kervella2004}. We adopted an interstellar reddening of $0.04 \pm 0.02$ mag to equalise the distance measurements in optical and infrared passbands. As a result, our best distance estimate, in the $K_s$ band, is $301.2 \pm 3.6 {\rm ~pc}$. This is slightly lower than the distance of $306.3 \pm 1.7 {\rm ~pc}$ from the inverse of the Gaia DR3 parallax\cite{gaiaedr3}.

\begin{table} \centering
\caption{\em Physical properties of \targ\ defined using the nominal solar units given by IAU 2015 Resolution B3 (ref.~\citenum{Prsa+16aj}). \label{tab:absdim}}
\begin{tabular}{lr@{~$\pm$~}lr@{~$\pm$~}l}
{\em Parameter}  & \multicolumn{2}{c}{\em Star A} & \multicolumn{2}{c}{\em Star B} \\[3pt]
Mass ratio $M_{\rm B}/M_{\rm A}$             & \multicolumn{4}{c}{$0.8892 \pm 0.0015$} \\
Semimajor axis of relative orbit (\Rsunnom)  & \multicolumn{4}{c}{$18.158 \pm 0.016$} \\
Mass (\Msunnom)                              & 1.4109 & 0.0035 & 1.2545 & 0.0037 \\
Radius (\Rsunnom)                            & 1.9901 & 0.0040 & 1.4599 & 0.0038 \\
Surface gravity ($\log$[cgs])                & 3.9898 & 0.0016 & 4.2079 & 0.0022 \\
Density ($\!\!$\rhosun)                      & 0.1790 & 0.0010 & 0.4032 & 0.0030 \\
Synchronous rotational velocity ($\!\!$\kms) & 18.340 & 0.036  & 13.454 & 0.036  \\
Effective temperature (K)                    & 6265   & 85     & 6320   & 90     \\
Luminosity $\log(L/\Lsunnom)$                & 0.740  & 0.024  & 0.486  & 0.025  \\
$M_{\rm bol}$ (mag)                          & 2.890  & 0.059  & 3.524  & 0.062 \\
Interstellar reddening \EBV\ (mag)           & \multicolumn{4}{c}{$0.04 \pm 0.02$}	\\
Distance (pc)                                & \multicolumn{4}{c}{$301.2 \pm 3.6$}\\[3pt]
\end{tabular}
\end{table}

\section*{Comparison with theoretical models}

We compared the measured properties of \targ\ to theoretical predictions from the {\sc parsec} 1.2 stellar evolutionary models \cite{Bressan+12mn}. We initially set the fractional metal abundance by mass to be $Z=0.014$ to match the mildly sub-solar metallicity found by CL10 in their spectroscopic chemical abundance analysis. In this case, an age of $2500 \pm 50$~Myr fits the masses and radii of the stars, but both are significantly cooler than the models predict. Increasing the $Z$ improves the match with the temperatures of the stars: for $Z=0.017$ we obtain good agreement in mass, radius and \Teff\ for an age of $2650 \pm 50$~Myr.

\begin{figure}[t] \centering \includegraphics[width=\textwidth]{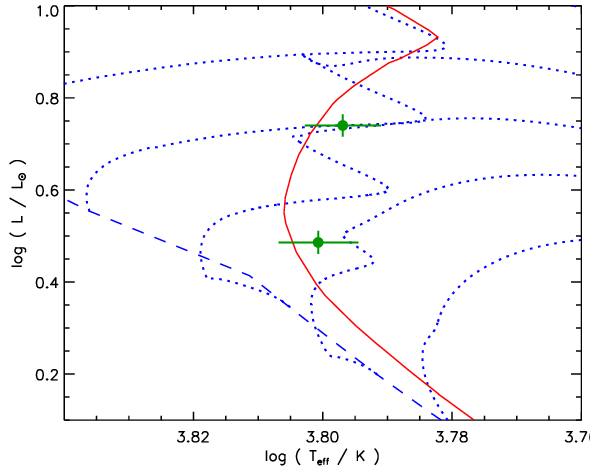} \\
\caption{\label{fig:hrd} Hertzsprung-Russell diagram for the components of \targ\ 
(filled green circles) and the predictions of the {\sc parsec} 1.2 models 
\cite{Bressan+12mn}. The dashed blue line shows the zero-age main sequence for 
$Z=0.0170$. The dotted blue lines show evolutionary tracks for $Z=0.017$ and masses 
of 1.1\Msun\ to 1.5\Msun\ in steps of 0.1\Msun\ (from bottom-right to top-left). 
The solid red line shows an isochrone for $Z=0.017$ and an age of 2650~Myr; it is not a perfect match in this diagram because it was chosen as the best fit to the masses and radii of the components of \targ.} \end{figure}

In Fig.~\ref{fig:hrd} we compare the properties of \targ\ to the predictions of the \textsc{parsec} models in a Hertzsprung-Russell diagram. This shows that both components are on the main sequence, although star~A is nearing the end of its main-sequence lifetime.

\section*{Summary and Conclusions}

\targ\ is a dEB containing two late-F type stars in a slightly eccentric orbit of period 5.49 days. We used the light curves from TESS Sector 57 and RV data from Popper and CL10 to measure the mass and radii of the companions. We measured the distance to the system using published \Teff\ values and surface brightness calibrations, finding a value close to but slightly shorter than the \gaia\ DR3 parallax distance.

The physical properties of the companions match theoretical models for a slightly super-solar metallicity and an age of 2650~Myr. This conflicts with the spectroscopic metallicity measurement of $\FeH = -0.12 \pm 0.07$ given by CL10.

Since there were only three mid-eclipse time measurements available from the TESS data, we also collected mid-eclipse times from previous literature works or open databases such as VarAstro portal. The eclipse times have an excess scatter around our fitted ephemeris. We conclude that further times of eclipse should be obtained to refine the ephemeris and measure its apsidal motion period.

\section*{Acknowledgements}

This research makes use of data obtained with the \textit{TESS} mission and accessed via the MAST archive at the Space Telescope Science Institute (STScI). The \textit{TESS} mission is supported by NASA’s Science Mission Directorate. STScI is operated by the Association of Universities for Research in Astronomy, Inc., under NASA contract NAS~5--26555. This study further makes use of data from the \textit{European Space Agency (ESA) mission Gaia}\footnote{\texttt{https://www.cosmos.esa.int/gaia}}, processed by the Gaia Data Processing and Analysis Consortium (DPAC\footnote{\texttt{https://www.cosmos.esa.int/web/gaia/dpac/consortium}}), with funding provided by national institutions, particularly those participating in the Gaia Multilateral Agreement. Additional resources employed in this work include the NASA Astrophysics Data System, the SIMBAD database operated at CDS, Strasbourg, France, and the \textit{arXiv} scientific preprint server maintained by Cornell University. ACK gratefully acknowledges the financial support of UK
Science and Technology Facilities Council (STFC) and the Faculty
of Natural Sciences, Keele University in the form of a PhD studentship. JS acknowledges support from STFC under grant number ST/Y002563/1.

\end{document}